\documentclass[10pt]{Fundamental_Dynamic_Units_Sauro}

\usepackage{cite} 
\usepackage{url}  
\usepackage{ifthen}  
\usepackage{multicol}   
\usepackage[utf8]{inputenc} 
\usepackage{graphicx}
\usepackage{setspace}
\usepackage{lscape}

\usepackage{appendix}

\newboolean{publ}

\newenvironment{bmcformat}{\begin{raggedright}\baselineskip20pt\sloppy\setboolean{publ}{false}}{\end{raggedright}\baselineskip20pt\sloppy}

\begin{document}
\begin{bmcformat}


\title{Fundamental Dynamic Units: Feedforward Networks and Adjustable Gates}


\author{Herbert Sauro\correspondingauthor%
         \email{Herbert Sauro - hsauro@u.washington.edu}%
         \hspace{2pt} and \hspace{1pt} Song Yang%
         \email{Song Yang - soyang@u.washington.edu}%
      }


\address{%
    Department of Bioengineering, University of Washington, Seattle, WA 98195, USA
}%

\maketitle


\begin{abstract}
        \paragraph*{Background:}
        The activation/repression of a given gene is typically regulated by multiple transcription factors (TFs)
that bind at the gene regulatory region and recruit RNA polymerase (RNAP). The interactions between the promoter
region and TFs and between different TFs specify the dynamic responses of the gene under different physiological
conditions.

        \paragraph*{Results:}
        By choosing specific regulatory interactions with up to three
transcription factors, we designed several functional motifs,
each of which is shown to perform a certain function
and can be integrated into larger networks. We analyzed three kinds of networks:
(i) Motifs derived from incoherent feedforward motifs, which behave as
``amplitude filters'', or ``concentration detectors''. These motifs respond maximally
to input transcription factors with concentrations within a certain range. From these motifs
homeostatic and pulse generating networks are derived. (ii) Tunable network motifs,
which can behave as oscillators or switches for low and high concentrations
of an input transcription factor, respectively. (iii) Transcription factor
controlled adjustable gates, which switch between {\bf AND/OR} gate characteristics,
depending on the concentration of the input transcription factor.

        \paragraph*{Conclusions:}
        This study has demonstrated the utility of feedforward networks and the flexibility of specific transcriptional binding kinetics
in generating new novel behaviors. The flexibility of feedforward networks as dynamic units may explain the apparent frequency that such motifs are found in real biological networks.

\end{abstract}

\ifthenelse{\boolean{publ}}{\begin{multicols}{2}}{}


\section*{Background}
In the last decade or so there have been significant advances in developing new
high throughput technologies such as microarray, ChiP-Chip or ChiP-seq~\cite{lee2002}
to uncover large scale cellular networks. Understanding these
networks has been facilitated by decomposing these networks into smaller
parts, so-called motifs, with known structure and function~\cite{milo2002}.
Two major strategies currently exist for accomplishing this task. The
first one, so called top-down analysis involves mining databases
for recurring patterns of interacting genes~\cite{milo2002, shen2002, yeger2004}, or in the
case of signaling networks, proteins-protein interactions~\cite{gerstein2006}. The emerging motifs are
analyzed for function, especially the specific roles they might serve in
a particular network. One interesting result from this work is the discovery of an enrichment in
so-called feedforward motifs~\cite{mangan2003a}, Fig. 1 (a), a three gene network.
Alon and colleagues~\cite{mangan2003a,mangan2003b,kalir2005,zaslaver2006}
subsequently demonstrated the functionality of these motifs using both
theory and experiment.

Another approach to understanding biological networks is the
bottom-up approach~\cite{guido2006}. Here knowledge of specific
protein-DNA interactions and signaling proteins has allowed
researchers to synthetically design gene regulatory networks. The
design of these networks has resulted in the construction of a wide
variety of network motifs such as oscillators, switches, and logic
gates~\cite{andrianantoandro2006, kaernweissbook, entus2007}. One of the goals
of synthetic biology is to design simple networks that can behave
as modules. In recent years
there has been a resurgence of interest in the quantitative
properties of gene regulatory networks~\cite{bolouri2002,kaern2003,alon2007}.
In addition, the ability to carry out relatively precise
measurements of gene activity using synthetic biology
techniques~\cite{andrianantoandro2006} and single cell
measurements~\cite{lahav2004,gevazatorsky2006} has led to considerable progress
in this field.

The top-down and bottom-up approaches are two ways to generate
functional modules. Based on a few network
motifs motivated by bottom-up approaches, we generated several novel functional network
motifs by exploring additional transcriptional control mechanisms and combinations of feedforward
networks.

One of the principles of gene regulation is regulated recruitment,
where transcription factors bind to the promoter regions of a gene,
recruiting other transcription factors and RNA Polymerase
(RNAP)~\cite{ptashne}. Furthermore, signaling molecules often
activate some of these transcription factors through binding
or phosphorylation. 
There are multiple ways by which these interactions
take place, such as blocking the promoter region, twisting
the DNA, Histone modification, and DNA looping, {\it
etc}.~\cite{bintu2005a}. In addition, the combinatorial complexity
of multiple transcription factors that bind to the same promoter region
will also determine the transcription rate ~\cite{buchler2003}.
Given all this complexity, a very useful approach to model these
kinds of interactions in gene regulatory networks is the
Shea-Ackers approach~\cite{sheaackers1982}. In this work, we used this method to
construct several different kinds of network motifs.

The network motifs we will discuss can be broadly classified into
three categories. (1) Feedforward related motifs:
obtained by additional feedbacks on the basic incoherent (type I)
feedforward network \cite{milo2002}. (2) Tunable motif: This
subnetwork can either behave as an oscillator or a bistable switch
depending upon the concentration of a transcription factor. (3)
Adjustable gates: This motif can switch between an AND and an OR gate
depending upon the concentration of a transcription factor. All
three network motifs arise from multiple transcription factor regulation
at a particular gene. In designing these {\it in silico} networks,
we didn't concern ourselves very much with exact mechanism such as DNA
looping {\it etc.}, but rather focused on the architecture of the resulting
network.

In the next section, we briefly review the Shea-Ackers approach, which is used
to compute the transcription rate of genes. Then, we will describe three different kinds
of models, amplitude filters,
tunable motifs, and adjustable gates. The latter two behave functionally
very differently depending upon the concentration of an input transcription
factor. A discussion and summary can be found in the last section.

\section*{Results and Discussion}

\subsection*{Modeling Gene Regulatory Networks Using Shea-Ackers Method}

A common method for modeling protein-gene interactions is by using
Michaelis-Menten or Hill function kinetics. However, in
the models we will describe in this paper, all kinetics will be
based on the Shea-Ackers
formalism\cite{sheaackers1982,bintu2005b}. This method estimates
the probabilities for different transcription states from
which an overall rate of transcription is derived. The approach is
less empirical and more flexible than the standard methods
and enables one to easily incorporate repressers and activators
into the equation. In addition the formalism has a relatively
straightforward relationship to the stochastic representations of
gene expression models. Finally, the free energy terms in the
formalism can be easily changed by altering the promoter region
which enables the models based on Shea-Ackers to be tested
experimentally.

We assume that the occupancy of the binding sites on promoters is
governed by equilibrium statistical thermodynamics
probabilities~\cite{sheaackers1982,bintu2005b,hill1960}. The
probabilities arise from the binding free energies, the free
energies of interaction between transcription factors bound to
adjacent sites, and the concentrations of the participating
transcription factors and the RNA polymerase. This assumption
holds when binding and unbinding
to the promoter is rapid~\cite{bintu2005b}. Slow binding
has been shown to result in stochasticity due to operator
occupancy fluctuations~\cite{kepler2001, hornos2005}, whereas for
small concentrations, the individual birth and death of proteins
introduce noise into the system~\cite{kaern2005}.

We illustrate the Shea-Ackers method with a given
gene regulated by two transcription factors, $T_1$ and $T_2$.
We assume that these three proteins bind at three distinct sites on
the promoter, providing nine possible combinations of binding. Each combination
is associated with a free energy that
is proportional to the probability of this state
and can be represented by the following equation ~\cite{bintu2005b}:
\begin{equation}
f = \exp (-\Delta G_f/{k_B T})\ {[T_1]}^{n_1} {[T_2]}^{n_2} [R]
\end{equation}
where $\Delta G_f$ is the free energy difference of the
bound and unbound states, and the factors $[T_1]$ and [$T_2]$ are
the concentrations of transcription factors, $R$ is the RNAP
concentration and $n_1$ and $n_2$ are the number of monomers
which combine into higher order multimers. $k_B$ and T are the
Boltzmann constant and absolute temperature respectively. From
Eq (1) the normalized probability of a given state is then
given by the ratio
\begin{equation}
  Z_f = \frac{\exp (-\Delta G_f/{k_B T}) {[T_1]}^{n_1} {[T_2]}^{n_2} [R]}{\sum_1^{9} \exp (-\delta G_f/{k_B T}) \ {[T_1]}^{n_1} {[T_2]}^{n_2} [R]}
\end{equation}
If we assume that the rate of transcription is proportional to the
relative probability when the polymerase is bound to the gene, then
we can partition the states into the
polymerase bound state ($Z_{on}$) and the
polymerase unbound state ($Z_{off}$)~\cite{buchler2003}. The probability of gene expression is then:
\begin{equation}
P=\frac{Z_{\mbox{\scriptsize on}}}{(Z_{on}+Z_{\mbox{\scriptsize off}})}.
\end{equation}
If a transcription factor is an activator, its free energy to
bind with the RNAP will be very low, and hence this
interaction will be favored. Whereas for an inhibitor, it would
be highly improbable for RNAP to be recruited for transcription.
Therefore, the interactions between the transcription factors and RNAP
determine the regulatory rules, which we will explore in the next
section for several different kinds of dynamical networks.

\subsection*{Feedforward Related Motifs}

First we assume that the
dynamics lumps together transcription and translation into one
process. Although explicit modeling and experiment has been shown
to give rise to interesting effects such as protein
bursts~\cite{mcadams1998, kaern2005} or oscillations as a result of transciption delay~\cite{Hasty2008}, we believe that the main
features of our models are captured by protein-DNA interactions.

\subsubsection*{Incoherent Type I Feedforward Related Networks}


High throughput approaches have uncovered several re-occurring
motifs termed the feedforward motif~\cite{milo2002,shen2002}.
Fig. 1(a) shows a common transcription
factor for two genes, where the third gene is regulated in a
feedforward fashion.
Fig. 1(b) shows a
protein-protein interaction in addition to this scheme.
%
%
Feedforward motifs can lead to two types of dynamics depending on
the nature of the two signals that converge on the third
gene. Fig. 2 illustrates a functional
representation of the schematic shown in
Fig. 1(a)~\cite{mangan2003a, mangan2003b}.
In this representation there are three different kinds of
interactions: protein degradation, protein synthesis and gene
regulation. In Fig. 2 the input transcription
factor $p_1$ modulates the activity of a target gene (G2)
directly and indirectly through the gene product $p_2$ of another
gene (G1), which is also a transcription factor. The
interaction between $p_1$ and $p_2$ at the promoter region of G2
and their ability to recruit RNAP determines the rate of
transcription of Gene G2. This type of architecture has been
shown~\cite{mangan2003b,kalir2005,zaslaver2006} to lead to two types
of dynamics depending on the nature of the regulation that occurs
at the target gene G2. If $p_1$, $p_2$ are both activators, the
gene circuit acts as a low pass filter, i.e.\ it is able to filter
out transient signals and transcribe only when the input signal is
long lived~\cite{mangan2003a}. If $p_1$, $p_2$ regulate G2 as an
activator and repressor respectively then the system can act as a
bandpass filter, since the delayed response of $p_1$ through $p_2$
tends to suppress activity of G2~\cite{mangan2003b}. This has been
suggested to be a general mechanism for speeding up response times in
transcriptional networks~\cite{mangan2003a, zaslaver2006}. Recently,
it has also been argued that the steady state characteristics of
such incoherent feedforward loops could be very important in
establishing spatial stripes and pulsed temporal expression profiles
of transcription factors involved in developmental
processes~\cite{ishihara2005}.


\subsubsection*{Simple Feedforward}

We assume that $p_1$ activates G1 and G2, $p_2$, the
gene product of G1, is recruited by $p_1$, and the protein complex
$p_1 p_2$ acts as a repressor of G2 (we assume that $p_2$ cannot
bind to G2 by itself). These assumptions lead to the following
rate equations,
\begin{eqnarray}
\label{simpleFeedforwardEquation}
\frac {d [p_2]}{dt}&=& \frac{a_1 + a_2 [p_1] }{1+ a_3 + a_4 [p_1]}-\gamma_1 [p_2] \\ \nonumber
\frac {d [p_3]}{dt}&=&\frac{b_1 + b_2 [p_1] }{1+ b_3 + b_4 [p_1] + b_5 [p_1][p_2]}-\gamma_2 [p_3],
\end{eqnarray}
where the transcriptional rates, and the ``lumped'' parameters are
derived in the appendix. $a_1, b_1$ represent leaky
transcription, which is due to the probability that RNAP can bind to
the operator region of the gene in the absence of a recruiting
transcription factor. At steady state, before G1 is saturated with
$p_1$, $p_2$ is proportional to $p_1$. The transcription rate
for G2 (assuming negligible leaky transcription) can be approximated
as,
\begin{equation}
T \propto \frac{{b_1}^{\prime} [p_1] }{1+ {b_2}^{\prime} [p_1]+
{b_3}^{\prime} {[p_1]}^2}
\end{equation}
The transcriptional rate rises in proportion to $p_1$ for small $p_1$, and
falls in proportion to $1/p_1$ for large values of $p_1$. Hence it
reaches a maximum at some intermediate value of $p_1$.


In Fig. 3, we show the steady state values of
$p_2$ and $p_3$ with respect to the input $p_1$. We note that the
concentration of $p_3$ has a maximum for a given value of $p_1$.

At low input concentration, $p_1$ transcribes G2 and G1, and hence
as its input level increases, $p_2$ tends to grow. Recruitment of
$p_2$ by $p_1$ at G2 makes it possible for the $p_1 p_2$ complex
to halt further transcription of G2. This module is aptly named an
``amplitude filter'' (originally called a ``band detector'',
\cite{basu2004, basu2005, kaernweissbook}, since its output is
maximal for a specific range of input. Such biphasic response has
also been discussed in other systems~\cite{wolf2003, mayya2005}.
Recently, Ishihara et. al.~\cite{ishihara2005} discussed the band properties of
such networks and used this to explain pulsed behavior and
patterning in {\it Drosophila} developmental processes.

\textsl{\\The effect of dimerization} \pb

In this section we will consider the effect of dimerization of
$p_2$, before it binds to the regulatory region of G2.
Dimerization has been shown to be important to
reduce the effects of stochastic fluctuations~\cite{bundschuh2003}
in a feedforward regulatory scheme. Here we discuss the steady
state behavior of a network regulated by $p_1$, which
recruits a dimer of $p_2$, and this complex is a repressor at G2.
The only changes that need to be made to
Eq.~\ref{simpleFeedforwardEquation} are to include the
dimerization equations. These modified equations are,

\begin{eqnarray}
\frac {d [p_2]}{dt}&=& \frac{e_1 + e_2 [p_1] }{1+ e_3 + e_4 [p_1]} -\gamma_1 [p_2] -2 (k_{d1} {[p_2]}^2-k_{d2} [p_d]) \\ \nonumber
\frac {d [p_3]}{dt}&=&\frac{d_1 + d_2 p_1 }{1+ d_3 + d_4 p_1 +  d_5 p_1 p_d}-\gamma_2 p_3 \\ \nonumber
\frac {d p_d}{dt}&=& k_{d1} {[p_2]}^2 -k_{d2} [p_d] -k_{d3} [p_d] \\ \nonumber
\end{eqnarray}

where in addition to the formation and dissociation of the dimer
complex, the dimer can also degrade. With these equations it is
easy to compute the steady state values of $p_3$ with respect to
the input $p_1$. In Fig. 4,
we plot the steady state value of $p_3$ as a function of
$p_1$ in the upper panel, which shows a much steeper fall off of the amplitude filter
curve, compared to the simple feedforward case discussed earlier.
The amplitude filter in this case, has a narrower bandwidth, and
hence its filtering capabilities of the input transcription factor
$p_1$ are much more enhanced. The sharp fall off is due to the
quadratic dependence of the amount of dimer with respect to the
input; as the input $p_1$ increases the amount of available dimer
$p_d$ to be recruited by $p_1$ at G2 increases, thus increasing
the amount of repression at G2. Dimerization of $p_1$ has the
effect of increasing both the width and shifting the peak of the
amplitude filter curve.


\textsl{\\Effects of mutations at G1 and G2} \pb
We now discuss the effects of two types of mutations at the binding
sites at G1 and G2, for the simple feedforward with dimerization of
$p_2$. Similar qualitative results can be obtained for the simple
feedforward and mixed feedforward. In general, a mutation at the binding
site tends to change the free energy of binding of the transcription factor,
generally decreasing the binding affinity. In
Fig. 4(a), a mutation at G1 reduces the ability of $p_1$ to bind to it. Hence larger
amounts of $p_1$ are required to achieve the same transcription rate
thereby shifting the amplitude filter peak to the right. In
Fig. 4(b), a mutation at G2 reduces the ability of $p_2$ to be recruited to G2.
This leads to a slower fall off, since repression does not
take place very efficiently. The two types of mutations can be used
to engineer the shape of the amplitude filter, by changing its
bandwidth and peak value~\cite{entus2007}.

\subsubsection*{Mixed Feedforward Motifs}

In Fig. 5, we show a slightly different
feedforward scheme, which is a functional interpretation of
Fig. 1b. The gene product $p_2$ is
shown as an activator of G2, whereas $p_1$ is a repressor of G2.
Also indicated in the figure is a protein-protein interaction,
whereby the input $p_1$ binds to $p_2$, and targets it for
degradation.

%
%

We assume that $p_1$ activates G1 leading to the production of
$p_2$; $p_2$ can individually bind to G2 and act as an activator.
Furthermore $p_1$ can be recruited by $p_2$ at the operator region
of G2, and together this complex acts as a repressor. $p_1$ binds
to $p_2$ and actively degrades it. One possible mechanism by which this can
occur is if $p_1$ labels $p_2$ with ubiquitin molecules for
proteolytic degradation~\cite{alberts}. The above assumptions lead
to the following rate equations,\
\begin{eqnarray}
\frac {d [p_2]}{dt}&=& \frac{c_1 + c_2 [p_1] }{1+ c_3 + c_4
[p_1]}-\gamma_c [p_1] [p_2] -\gamma_1 [p_2] \\ \nonumber \frac {d
[p_3]}{dt}&=&\frac{d_1 + d_2 [p_2] }{1+ d_3 + d_4 [p_1] + d_5
[p_2] + d_6 [p_1] [p_2]}-\gamma_2 [p_3],
\end{eqnarray}
where the transcriptional rates, and the ``lumped'' parameters are
derived in the appendix. In the equation for $p_2$, the extra
degradation term is due to the protein-protein interaction between
$p_1$ and $p_2$. For this system of equations the plots shown in
Fig. 6 show the behavior of the
steady state value of $p_3$, demonstrating the amplitude filter
effect.


As $p_1$ increases, $p_2$ begins to grow and transcribe G2;
however, two key factors prevent G2 from being continually
transcribed with further increases in $p_1$. The first is that
$p_2$ is targeted by $p_1$ for degradation, and the second is that
$p_2$ binds to G2 and recruits $p_1$, which turns off the
transcription.

In both the simple incoherent Type I feedforward, with/without
dimerization, as well as the mixed feedforward, we have shown how
it is possible to obtain a basic amplitude filter. The property of
such an amplitude filter can now be further exploited to generate
new types of networks when additional feedbacks are added to this
basic motif.

\textsl{\\Application 1: Time Ordering}\pb

The amplitude filter or ''concentration detector'' has been shown
to perform temporal processing functions such as pulse
generation~\cite{basu2004}. Pulse-like behavior has been simulated
in~\cite{ishihara2005}, with a series of cascaded feedforward
loops, which in fact use the amplitude filter property of the
networks to generate pulsatile behavior. We use a similar idea
where a single transcription factor could serve as an
input to several amplitude filters, each of which has a different
shape. In particular we assume that for each amplitude filter
module, the peaks of $p_3$ occur at different
input transcription factor ($p_{1opt}$) concentration values.
Then, as the input concentration crosses $p_{1opt}$, the amplitude filters get
activated in a sequence, depending on how far apart $p_{1opt}$ are
for different amplitude filters~\cite{mangan2003b}.


A single input can therefore activate several genes in a sequence. One
example could be the sequential release of different
proteins required in a certain developmental process.

Fig. 8a plots $p_3$ as a function of the
input transcription factor concentration, showing a shifted peak for
the two amplitude filters with respect to the same input
transcription factor. In Fig. 8b, the
temporal profile for the outputs are shown assuming that the input
concentration is ramped up as a function of
time. This leads to an ordered protein production in time.

 Kashtan et. al.~\cite{kashtan2004}, have explored the
consequences of multi-output networks regulated by
feedforward networks. They show through simulations that in some of
these cases feedforward loops with a common input can regulate
genes in a temporal order. Such temporal order can occur in
multi-output feedforward loop systems such as in the {\it E. coli}
flagellar synthesis regulation systems, where proteins need to
assembled in a timed fashion to make up the flagellar basal-body
motor~\cite{kalir2001}. It has also been found
in~\cite{eichenberger2004}, that the logic of the program of gene
transcription during differentiation in {\it Bacillus Subtilis}
sporulation involves a series of feedforward loops that generate
gene transcription in a pulse like manner.

\textsl{\\Application 2: Homeostatic Networks}\pb

Homeostatic networks are important in several biological systems.
One example in which homeostasis has been shown to occur through
integral feedback control is in the chemotaxis network in {\it E.
coli} (\cite{yidoyle2000}). Our motivation was to design a network
using the amplitude filter which would display homeostasis to
input perturbations.

Using the output of the amplitude filter module as an input to
itself, one can construct the motif shown in
Fig. 9(a), where the filter module's output
$p_3$ participates as a transcription factor for gene G1, whose
protein product $p_1$ serves as the input
to the filter module. The feedback could be positive or negative, depending on whether the
amplitude filter output $p_3$ is an activator or repressor of G1.
In addition, an external input protein A can bind as an
activator to gene G1. Depending on the interaction between
the input control A, the feedback $p_3$, and RNAP, we obtain
different types of behavior. Consider $p_3$ as an activator of
G1 in Fig. 9(a), we assume that the
interaction between $A$, $p_3$ and RNAP is such that we obtain an
{\bf AND} gate at G1, {\it i.e.} G1 expresses only when both $A$
and $p_3$ are present. The equation that describes the
additional variable, $p_1$, is given by
\begin{eqnarray}
\frac{d p_1}{d t}&=&\frac{u_1+u_2 [A][p_3]}{1+u_3 + u_4 [A]+u_5
[p_3]+u_6 [A][p_3]}-\gamma_3 p_1
\end{eqnarray}
which is the equation for the input of the amplitude filter, and the
equation for $p_3$, the output of the amplitude filter, is the same as in
Eq 6. The steady state plot for the output $p_3$ as a function of the
input A, is shown in Fig. 10(a).



Figure 10 shows that the steady state values of
$p_3$ are constant even for large inputs. As the input A of
the filter module increases, the output $p_3$ decreases (assuming
that $p_3$ is maximal at the initial value of A). This is because
the input transcription factor concentration moves away from
$p_{1opt}$, at which the maximal value of transcription occurs.
Therefore, this decreases the transcription of the filter module,
and since its output feeds back as an activator to G1, its input
level tends to decrease the transcription of G1. Essentially the
filter module balances the increase in the input A to G1 by
decreasing its output. As seen in the lower plot of
Fig. 10(a), $p_3$ stabilizes to an almost
fixed value even though the input control $A$ increases in time.
This is an example of a homeostatic gene network which fixes its
response to input transcription factor concentrations. The other
case shown in the scheme in Fig. 10(b),
where $p_3$ is a repressor for G1.
In the regulation at G1, A as an activator and $p_3$ is a repressor. The
negative feedback of $p_3$ into G1 suppresses the input to the
amplitude filter. If A decreases, the input to the
filter module $p_1$ also decreases, this reduces the transcription rate
(assuming that at steady state the value of $p_3$ is maximal).
This then lifts the repression from G1, and the input to the
filter module increases, thereby balancing the effect due to the
reduced input A, and achieving the same steady state as before.
The above two circuits produce a fixed amount of output proteins,
even though the input might vary by a large amount. In the first
case it is homeostatic to an increase in the concentration value
of the input transcription factor, in the second case it is
homeostatic with respect to a decrease in the input concentration.

\subsection*{Tunable Motifs}


In this section we discuss an example of a gene network that
exhibits oscillatory dynamics at small values of an input
transcription factor, and bistability for high input values.
This network therefore implements two
different types of motif functionality, depending upon the
concentration of an input transcription factor. Recently Voigt et.
al.~\cite{voigt2005} discussed a model of a network in the
Bacillus sporulation pathway,  which was shown to exhibit one of
two alternatives, i.e.\ either a bistable switch or oscillatory
behavior, depending on the environmental conditions.  A
synthetically constructed network exhibiting multifunctionality in
{\it E. coli}~\cite{atkinson2003} was demonstrated to be able to flip
function from an oscillator to a switch, by removing a particular
interaction.

There are now several examples of synthetically designed genetic
networks such as switches~\cite{gardner2000,becskei2001} and
oscillators~\cite{elowitz2000,andrianantoandro2006} that use the
common rules of mutual inhibition and activation to achieve a
desired functional behavior. The network we will describe is
motivated by the work of Gardner et. al.~\cite{gardner2000}, where
the authors designed a toggle switch. In ~\cite{hasty2001}, it was
discussed how a toggle switch could be converted into a relaxation
oscillator by suitably manipulating the basic toggle switch
network by adding extra regulation. We further extend this design
by introducing a new type of regulation, which involves an
external transcription factor whose concentration can flip the
system function between an oscillator and a switch.

Fig. 11 shows two mutually repressing genes, G1 and G2.
The repression is assumed to occur through
tetramer binding of each of their gene products, $p_1$ and $p_2$,
the gene products of G1 and G2, ($p_1$ binds to
G2, and $p_2$ binds to G1). Hence, if these were the only
interacting genes in the network, the system could be in one
of two stable states, i.e., G1 is fully expressed, and G2 is
silent or {\it vice versa}. Gene G3 is activated by $p_2$, and
its product $p_3$ further activates G1. The feedback of $p_3$ to
G1 has the effect of turning the bistable switch into a relaxation
oscillator~\cite{hasty2001}. Assuming that initially G2 is {\bf ON},
which causes G3 to get transcribed and $p_3$ to grow, $p_3$ then
activates gene G1, leading to the growth of $p_1$. Thus G1
switches to a high state, which ultimately shuts down gene G2 due
to its repressive effects. This leads to a decrease in $p_3$ which
subsequently turns G1 {\bf OFF}, thereby completing one cycle of
relaxation oscillations.

Consider now an extra piece of regulation, an external input $A$,
as an activator of gene G1. $A$ can bind to the promoter region and
activate G1, but can also cooperatively bind with the tetramer of
$p_2$, which represses G1. $p_3$ can also bind cooperatively
with the tetramer of $p_2$, which has the effect of repressing G1.
The two activators $A$ and $p_3$, however, are assumed to be
mutually exclusive, i.e.\ both $A$, and $p_3$ cannot bind
together, but each can individually bind to the DNA. As described
in the Appendix, these regulatory rules make G1 behave like an {\bf
OR} gate with respect to the inputs $A$ and $p_3$. From the above
regulatory mechanisms, the following equations for the protein
dynamics emerge,

\begin{eqnarray}
\frac {dp_1}{dt}&=& \frac{m_1 [A]+m_2 [p_3]}{1+m_3 [A]+m_4
[p_2]^4+m_5 [A][p_2]^4+m_6 [p_3] +m_7 [A] [p_3]} -\gamma_1 p_1 \\ \nonumber
\frac {dp_2}{dt}&=&\frac{n_1}{1+n_2+n_3 [p_1]^4} -\gamma_2 p_2, \\ \nonumber
\frac{dp_3}{dt}&=&\frac{o_1+o_2 [p_2]}{1+ o_3 + o_4 [p_2]}
-\gamma_3 p_3,
\end{eqnarray}

We now consider approximating the transcription rate for G1 for
small and large values of the control transcription factor, $A$. For
small $A$,
\begin{equation}
T \simeq \frac{m_2 [p_3]}{1+m_4 [p_2]^4 + m_6 [p_3] +m_7 [p_2]^4
[p_3]}
\end{equation}
G1 is activated by $p_3$, and repressed by $p_2$. As
discussed earlier $p_3$ toggles the bistable switch
formed between the gene products of G1 and G2. For large $A$,
\begin{equation}
T \simeq \frac{1}{\frac{m_3}{m_1}+\frac{m_5}{m_1}[p_2]^4}
\end{equation}
which is the transcription rate, one would obtain a toggle
switch, between G1 and G2.  When $A$ is large, it is more likely
for $A$ to bind to the DNA than its competitor $p_3$ and the
system is put into the bistable regime. The input $A$ can
therefore be used to tune the system into a relaxation
oscillator, or a switch. Fig. 12 shows the
bifurcation plot for the steady state values of $p_1$ as a
function of the input transcription factor $A$. The plot shows a
subcritical Hopf bifurcation~\cite{strogatz,kuznetsov} for $p_1
\simeq 9$, and a saddle-node bifurcation at $p_1 \simeq 35$. As
discussed earlier, the system exhibits oscillations for $A< \simeq
11$ and bistability for $A> \simeq 35$. In the right panel of
Fig. 12, the time series plots for $p_1$,
$p_2$, $p_3$, are displayed, which show oscillations, for the case
where $A=1$. In the lower right panel of
Fig. 12 the bistable behavior of $p_1$ is
displayed (for the input $A=50$), taking one of two values,
depending on the initial conditions.


\subsection*{Adjustable Gates}

The transcriptional interaction between two transcription factors
and RNAP has previously been shown to generate several instances of
logic such as {\bf AND, OR, XOR}, {\it etc.}~\cite{buchler2003}.
Multiple transcription factors can regulate the gate properties of a
network through their input concentrations. Alon and
colleagues~\cite{setty2003} have studied the gate properties of the
regulation of the {\it lacZYA} operon in {\it E. coli}; and by using a
mathematical model, their work shows that the regulation can be made to behave
as a fuzzy AND, a pure AND, and an OR logic gate. In this section we
consider gene regulation by three transcription factors which have the
property of switching between two different logical functions,
depending on one of the input transcription factor concentrations.
Further we propose that such a circuit with appropriate feedback can
be made to exhibit other kinds of functionality such as homeostasis
or oscillatory behavior.

Consider gene G1 transcribed by the interaction of three
transcription factors, $p_1$, $p_2$ and $p_3$. $p_3$ is the control
that determines the output logic. We make the following
assumptions: binding of all three transcription factors and
RNAP to the gene is unlikely; individual binding of $p_1, p_2,
p_3$ and RNAP is also unfavorable; we also assume that
transcription is not leaky. With these assumptions, transcription
occurs due to the binding of the following complexes $p_1 p_2 P,
p_1 p_3 P, p_2 p_3 P$. The transcription rate, takes the following
form,
\begin{equation}
T=\frac{r_1 p_1 p_2 + r_2 p_1 p_3 +r_3 p_2 p_3}{1+ r_4 p_1+ r_5
p_2 + r_6 p_3  +r_7 p_1 p_2 + r_8 p_1 p_3 +r_9 p_2 p_3+r_{10} p_1
p_2 p_3}
\end{equation}
The above formula can be simplified for two cases, i.e.\ high
and low values of the concentration of the control transcription
factor, $p_3$. For low values of $p_3$,
\begin{equation}
T \simeq \frac{r_1 p_1 p_2}{1+ r_4 p_1+ r_5 p_2+r_7 p_1 p_2 }
\end{equation}
which implies that the transcription is activated only when both
$p_1$ and $p_2$ are present, implementing an {\bf AND}
gate. For high values of $p_3$, the transcription rate is,
\begin{equation}
T=\frac{{r_2}^{\prime} p_1  +{r_3}^{\prime}p_2}{1  +{r_8}^{\prime}
p_1 +{r_9}^{\prime} p_2 +{r_{10}}^{\prime} p_1 p_2}
\end{equation}
where ${r_i}^{\prime}=\frac{r_i}{r_6}$, for $i=2,3,8,9,10$. Here the
transcription is activated when either $p_1$ or $p_2$ is present,
implementing an {\bf OR} gate. The control $p_3$ is able to switch from an
{\bf AND } to {\bf OR} gate. In Fig. 13 we plot the
transcription rates as a function of the input transcription factors, $p_1$ and
$p_2$.


The ability to switch from one kind of logical function to another
by varying the control $p_3$ opens up the possibility to use such
a motif in a gene network with other interacting genes. Consider
Fig. 14(i), a regulatory circuit
with G1 having three inputs $p_1, p_2$ and the control $p_3$,
which emerges from a long negative feedback loop from its gene
product $p_5$ in the following way: The gene G2 constitutively
produces protein $p_4$, which forms a protein complex with $p_5$.
$p_4$ is also a transcription factor for the gene G3, which
produces protein $p_3$, and hence this is how a feedback into G1
is achieved. We further assume that $p_4$ binds to the regulatory
region of G3 as a tetramer. The transcription factors $p_1$ and
$p_2$ are assumed to be external to the system, and can be
either set to a constant or a time dependent value.
In our case we shall fix the value of $p_2$, but allow
$p_1$ to fall to low levels, starting at some fixed value.
However, the dynamics of the network is symmetric with respect to
$p_1$ and $p_2$, and hence we could have equally well chosen
to vary $p_2$ and keep $p_1$ fixed. Fig. 15 shows
the logical structure for the adjustable gate network. Note that
the output from the gate determines its behavior. The
differential equations for the rates of production of the various
species are,
\begin{eqnarray}
\frac{d[p_5]}{dt} &=& \frac{r_1 [p_1] [p_2] + r_2 [p_1] [p_3] +r_3
[p_2] [p_3]}{1+ r_4 [p_1]+ r_5 [p_2] + r_6 [p_3]  +r_7 [p_1] [p_2]
+ r_8 [p_1] [p_3] +r_9 [p_2]
[p_3]+r_{10} [p_1] [p_2] [p_3]} \\ \nonumber \\ \nonumber
&-& k_1 [p_4] [p_5] + k_2 [C] - \gamma_{p_5} [p_5] ,\\ \nonumber \\ \nonumber \frac{d[C]}{dt}
&=& k_1 [p_4] [p_5] - k_2 [C] ,\\ \nonumber \frac{d [p_4]}{dt} &=&
c_0 - k_1 [p_4] [p_5] + k_2 [C] -\gamma_{p_4} [p_4],\\ \nonumber
\frac{d [p_3]}{dt} &=&
\frac{v_2 [p_4] ^4}{1 + v_4 [p_4] ^4} - \gamma_{p_3} [p_3], \\
\nonumber
\end{eqnarray}
For gene G1, the input $p_2$ is held constant, but the input $p_1$
is made to decay from some initial value.
Gene G2 produces $p_4$ constitutively, and
is sequestered by the output of G1, {\it i.e.} $p_5$, into the
complex $C$. Since $p_4$ is a transcription factor for G3, its
sequestration away from G2 results in a lower value of the protein
$p_3$. Hence the control $p_3$ is changed, the gate properties
of G1 switch between {\bf AND}/{\bf OR}. The negative feedback
arising from G1 onto itself is inhibitory, due to the complex
formation. In Fig. 14(ii), the upper plot shows
the steady states of the value of the control $p_3$ as a function
of the input $p_1$. A supercritical Hopf bifurcation is seen to
occur at $p_1 \simeq 2.5$. Fig. 14(ii) lower
right-hand plot shows steady oscillations of $p_3$. The oscillations
arise due to the gate properties of G1. Initially, when the system
is at steady steady state, and the inputs $p_1$ and $p_2$ are fixed,
$p_3$ has a small value which makes G1 behave as an {\bf AND} gate.
since both inputs are present, the output $p_5$ is high. There
is thus considerable sequestration of $p_4$ due to the complex
formation. This implies a reduced production of $p_3$, which is
consistent with G1 being in the {\bf AND} state. Now if one of the
inputs to the system is removed ({\it e.g} by making $p_1$ decay),
since initially the gate is in the {\bf AND} state, the output $p_5$
decreases. This results in the release of $p_4$, which begins
increasing the transcription of G3; then $p_3$ increases, and the
gate G1 switches to the {\bf OR} state. But in this state, G1 can
be transcribed by $p_2$, and the output $p_5$ increases. This once
again results in sequestration of  $p_4$, and finally reduces
the control $p_3$, and in this way we complete one cycle. The system
therefore switches back and forth between the two states and hence
this leads to oscillations.


We now describe another application of an adjustable gate. However
in this case, the gate properties are reversed, i.e., the gate
implements an {\bf OR} gate for low input control transcription
factor concentration, and an {\bf AND} gate for high input control
transcription factor concentration. We discuss the nature of the
regulation and its consequences without simulations, since this
case is very similar to the previously described model. For this
case we assume the following regulatory rules: all three
transcription factors and RNAP bind to the gene;
each of the transcription factors $p_1, p_2$ and
RNAP bind individually to the gene; and the complex $p_1$ $p_2$ and
RNAP can bind to the gene. These assumptions lead to the following
rate law,
\begin{equation}
T=\frac{s_1 p_1 p_2+ s_2  p_1 +s_3  p_2 +s_4 p_1 p_2 p_3}{1+ s_5
p_1+ s_6  p_2 +s_7 p_3  +s_8 p_1 p_2 +s_9 p_1 p_3+s_{10} p_2
p_3+s_{11} p_1 p_2 p_3}
\end{equation}
By inspection it is clear that for low $p_3$, G1 behaves like an
{\bf OR} gate with respect to $p_1$ and $p_2$, whereas for high $p_3$, it
behaves like an {\bf AND} gate. If we now consider the same network
as described above but substitute this motif into G1, then the
system shows an almost homeostatic behavior with respect to a change
in its input. This is easy to understand since initially, when both
inputs are present, the output $p_5$ must be large. Due to the
nature of the feedback, this determines the value of $p_3$ to be at
a low level. The system is therefore initially in the {\bf OR}
state. If now one of the inputs is suddenly decreased, since the
system is initially in an {\bf OR} state, the output would continue
to be high. There would be a small transient due to the sudden
change in input, but the system would reach a steady state as
before. This then allows this network to be fairly unperturbed to
changes in one of its input transcription factors.


What could be the function of such networks with adjustable gates?
In the first case (with {\bf AND} gate properties for low $p_3$),
a sudden drop in a transcription factor concentration could set up
oscillatory patterns in the network which would then signal the
next program to be carried out by the genetic network. In the
second case (with the gate properties reversed), it is clearly
useful to have a homeostatic network which works in such a way so
as to counteract any sudden changes in input transcription factor
concentration levels.

Although we have not found explicit examples for many of the
networks we have discussed, which display such complex behavior,
we believe that such an endeavor is worth exploring.

\section*{Conclusions}

As a first step towards recognizing and understanding large
complicated pathways, we have discussed in this work the modular
design of several functional network motifs. Each of the networks
consists of genes which are regulated by multiple transcription
factors. The combinatorial regulation was explored in each case, and
the networks which emerged were found to have very distinct
properties. Our modeling procedure used the Shea-Ackers method~\cite{bintu2005b}, which allowed us to derived the rates of
transcription which were then used to explore the network dynamics.
The networks could broadly be classified into: networks which are
derived from incoherent feedforward motifs; and networks which can
change their gating properties based upon an external input.

We first discussed the steady state properties of feedforward
networks, which can be used as amplitude filters. We found that both
the Type I simple and mixed feedforward networks led to a similar
design, i.e.\ filtering out the input transcription factor
concentration, although both networks worked through different types
of interactions. Furthermore, we discussed the effects of
dimerization for a simple Type I feedforward, which has the effect
of narrowing the bandwidth of the amplitude filter. To study the
filter characteristics of these networks, we simulated the effects
of mutations which would change the protein-DNA binding strengths.
These generally have the effect of shifting the amplitude filter
curve and modulating its bandwidth. Furthermore we described how
these motifs can be applied to a biological setting. By having a
common transcription factor as the input to two amplitude filter
modules, but with shifted filter characteristics, it was possible to
obtain a time ordered response of protein production. Homeostatic
networks emerged if the output of the amplitude filter was fed back
to itself. This network was found to be resilient in its output to
either increasing or decreasing values of an external input
transcription factor, depending on whether the feedback was assumed
to be positive or negative respectively.

We next described a tunable motif network, where the regulation at
one of the genes made it possible for the networks to exhibit
bistability or oscillatory behavior if one of the external input
transcription factor concentration was made to increase/decrease
respectively. Finally we discussed gate properties of regulation at
a gene, which can be made to switch between an {\bf AND/OR}
depending on the one of the input transcription factor
concentrations. 

\section*{Methods}

\subsection*{Simulations}

All simulations were carried out using the Systems Biology Workbench
(SBW) tools~\cite{Sauro:Omics}: the network designer,
JDesigner, the simulation engine Jarnac~\cite{Sauro:2000}.
Bifurcation diagrams were computed using SBW with an interface to
MATLAB~\cite{Cameron:MATLAB}, and a bifurcation discovery
tool~\cite{ChickarmaneBifTool}. Bifurcation plots were also computed
and cross checked using
Oscill8~\footnote{http://sourceforge.net/projects/oscill8}, an
interactive bifurcation software package which is linked to
AUTO~\cite{Do81}, and SBW~\cite{Sauro:Omics}. In all our simulations
the species concentrations are regarded as dimensionless, whereas
the kinetic constants have dimensions of inverse time, with
dimensionless Michaelis-Menten constants. All models are available
as Jarnac scripts (supplied in the supplement) which can be easily translated to
SBML~\cite{hucka:2002d} using the JarnacLite tool that is part of
the SBW suite~\cite{Sauro:Omics}.

\section*{Author contributions}
    HS helped conceive and fund the project; SY contributed to manuscript preparation and final data analysis.

\section*{Acknowledgements}
  \ifthenelse{\boolean{publ}}{\small}{}
This work was supported by grants from the National Science Foundation (0432190 and FIBR 0527023) to HMS. The authors wish to thank Carsten Peterson for useful discussions and most significantly to Vijay Chickarmane for technical assistance in carrying out the simulations and deriving the transcription factor binding kinetics. VC declined to be a co-author on the manuscript but was also supported by NSF FIBR 0527023.


{\ifthenelse{\boolean{publ}}{\footnotesize}{\small}
 \bibliographystyle{Fundamental_Dynamic_Units_Sauro}  
  \bibliography{Fundamental_Dynamic_Units_Sauro} }     


\ifthenelse{\boolean{publ}}{\end{multicols}}{}



\section*{Figures}
  \subsection*{Figure 1 - Feedforward motifs extracted from analyzing}
      protein-protein and protein-DNA interaction databases. (a) Simple
      feedforward, the thin arrows indicate transcriptional activity, (b)
      Mixed feedforward; In addition to the transcriptional activity, the
      thick arrow indicates a protein interaction.

  \subsection*{Figure 2 - Interpretations of the feedforward motifs shown in Fig. 1(b) in terms of a reaction network.}
		$p_1$ is the input to the system comprising of genes G1 and G2. G1
		produces $p_2$, and G2, $p_3$, $p_1$ activates G2, $p_2$ represses
		G2. The line which ends with a dot represents activation, and the
		line which ends with a small line perpendicular to it represents
		repression.
		
  \subsection*{Figure 3 - Steady state values of $p_3$, and $p_2$ for the simple
incoherent Type I feedforward network as a function of the input
factor $p_1$}

  \subsection*{Figure 4 - Steady state value of $p_3$ as a function of $p_1$ for the
simple feedforward with dimerization, and for the consequences of
mutations at G1 and G2.}
      Upper plot: The effects of dimerization in a
simple feedforward, as compared to a curve obtained for a simple
feedforward network. Lower Plots: (a) $p_3$ as a function of $p_1$
for wild (strong binding of $p_1$ to G1) and mutant types (weak
binding of $p_1$ to G1). (b) $p_3$ as a function of $p_1$ for wild
(strong binding of $p_1 p_2$ to G2) and mutant types (weak binding
of $p_1 p_2$ to G2).

  \subsection*{Figure 5 - Time ordered production of proteins can be achieved by
feeding in a common input into two or more amplitude filter modules
that have different maximal responses.}

  \subsection*{Figure 6 - Sample figure title}
      (a): Steady state values of the amplitude filter outputs
as a function of the input A, showing the shifted output peaks with
respect to the input transcription factor concentration. (b): The
amplitude filter outputs as a function of time showing the time
ordering.

  \subsection*{Figure 7 - Homeostatic networks built out of the amplitude filter motifs.}
      (a) Feedback, $p_3$ from the output amplitude filter positively regulates
gene G1, (b) Feedback, $p_3$ from the output amplitude filter negatively
regulates gene G1.

  \subsection*{Figure 8 - Homeostatic networks built out of the amplitude filter motifs.}
      (a): Steady state values of $p_3$ as a function of the input A. (b): $p_3$,
$p_1$ and the input A as a function of time.

  \subsection*{Figure 9 - Interpretations of the second feedforward motifs shown in
Fig. 1 in terms of a reaction network.}
      $p_1$ degrades $p_2$ via a binding reaction, at G2, $p_2$ activates G2, whereas
$p_1$ acts as a repressor.

  \subsection*{Figure 10 - Steady State response of a mixed feedforward network as a
function of input factor $p_1$.}

  \subsection*{Figure 11 - A cartoon of the tunable motif which shows three
interacting genes: G1, G2 and G3, and their gene products $p_1,
p_2, p_3$ respectively.}
      Regulation at G1 includes positive
regulation by an external transcription factor A and $p_3$, and
repression by $p_2$. The transcription factor A for large
concentrations can out compete $p_3$ in binding to G1. The effect
of this is to break the feedback between the two antagonists which
make up the toggle switch, and hence the network exhibits
bistability for large A.

  \subsection*{Figure 12 - Bifurcation diagram and time-series plots for the tunable
motifs.}
      The left figure shows the bifurcation plot for the steady
state value of $p_1$, as a function of the transcription factor
concentration A. A subcritical bifurcation occurs at $A \sim 9$, and
oscillations occur for $A<=11$. At $A \sim 35$, a saddle-node
bifurcation occurs which gives rise to the appearance of two stable
steady states. Figure (a) shows time series plots for $p_1$, $p_2$,
$p_3$, which show oscillations, for the case where $A=1$. Figure (b)
shows the final steady state of $p_1$ reaching either of two values,
depending on the initial conditions. This is due to the bistable
behavior of the circuit. The parameter for this case is $A=50$

  \subsection*{Figure 13 - The figures show the transcription rate as a function of the
transcription factor concentrations, $p_1, p_2$, for fixed values of the
control transcription factor $p_3$.}
      (a), $p_3=1$, represents an {\bf AND} gate, (b), $p_3=10$, represents an {\bf OR} gate.

  \subsection*{Figure 14 - An application of adjustable gates}
      Panel (i) shows the schematic of the network while Panel (ii) shows the simulations
results. In (ii), the upper plot shows the bifurcation plot of
$p_3$, as a function of the input $p_1$. A supercritical Hopf
bifurcation is seen to occur at $p_1 \simeq 2.5$. The lower panel in
(ii)(a) shows the decay of $p_1$, while the lower panel in (ii)(b)
shows the system reaching steady oscillations of $p_3$, after an
initial transient.

  \subsection*{Figure 15 - A logical diagram of the schematic shown in
Fig. 14.}
      The triangle symbol represents a Boolean
inverter. The action of the Boolean gate switches the AND-OR gate
between AND and OR functions

\appendix

\section{Appendix 1}

In this section we demonstrate the equivalence of the
thermodynamic (Shea-Ackers) approach with that of the enzyme
kinetic approach to derive the equations for the feedforward
network. Although this is well known we provide this derivation
explicitly to show how a particular choice of interaction can be
used to determine a reaction scheme. The reaction scheme is
required if a stochastic simulation is to be performed. From Fig.
2 we see that $p_1$ is an activator for G1. RNAP can also bind to
G1 in the absence of $p_1$, but at a lower rate. As described in
the main text, such binding leads to ``leaky transcription''. At
the operator sites of G2, $p_1$ can bind and recruit RNAP to the
operator site of the gene. However, $p_1$ also recruits $p_2$, the
gene product of G1, and in this state, transcription is repressed.
We also assume that $p_2$ binding to G2, occurs only as
recruitment, and cannot occur by itself. From these assumptions,
we can draw up the following truth tables for the transcriptional
regulation at G1 and G2.

\begin{table}[htb]
\centering
\parbox[t]{0.45\textwidth}{
\centering
\begin{tabular}{ccr}
\hline
$p_1$ & $P$ & Rate \\
\hline \hline
0 & 0 & $1$ \\
1 & 0 & $\alpha_1 [p_1]$ \\
0 & 1 & $\alpha_2 [P]$ \\
1 & 1 & $\alpha_3 [p_1~P]$ \\
\hline
\end{tabular}
}
\parbox[t]{0.45\textwidth}{
\centering
\begin{tabular}{cccr}
\hline
$p_1$ & $p_2$ & $P$  & Rate \\
\hline \hline
0 & 0 & 0 & $1$ \\
0 & 1 & 0 & $\times$ \\
1 & 0 & 0 & $\beta_2 [p_1]$ \\
1 & 1 & 0 & $\beta_3 [p_1][p_2]$ \\
0 & 0 & 1 & $\beta_4 [P]$ \\
0 & 1 & 1 & $\times$\\
1 & 0 & 1 & $\beta_6 [p_1][P]$ \\
1 & 1 & 1 & $\times$ \\
\hline
\end{tabular}
} \caption{Logic for the transcriptional regulation underlying
simple feedforward genes. In the case where transcription is
unlikely to occur the rate is denoted by $\times$ and assigned to
zero in the calculations.}
\end{table}

In the above table and the subsequent tables the terms $\alpha_i,
\beta_i$, are related to the free energies of binding through
$\exp{\frac{\delta G}{k_B T}}$. The logic in Table 1 can be
translated to the rate of transcription by computing the
fractional probability of each of the genes G1 and G2 being bound
by RNAP. The transcription rates, $Tr_{G1}, Tr_{G2}$, are
proportional to the probability of occupancy of RNAP, which can be
computed from the above two tables to be,
\begin{eqnarray}
Tr_{G1}&\propto&\frac{\alpha_1~[P] +
\alpha_3~[p_1]~[P]}{1+\alpha_1~[P]+\alpha_2~[p_1]+\alpha_3~[p_1]~[P]}\\
\nonumber Tr_{G2}&\propto&\frac{\beta_4~[P] +
\beta_6~[p_1]~[P]}{1+\beta_4~[P]+\beta_2~[p_1]+\beta_3~[p_1]~[p_2]+\beta_6~[p_1]~[P]}\\
\nonumber
\end{eqnarray}
The transcription rates are then used in the rate laws described
in Eq. 4. We now compute the above result using the reaction
kinetics approach. Assume that for the gene G1, there are 4
states: G1-unbound or free, $G1p_1$-bound by $p_1$, $G1P$-bound by
RNAP, $G_1p_1P$-bound by $p_1$ and RNAP. Then it follows that,
\begin{equation}
[G1] + [G1p_1] +[G1P] + [G1p_1P] =1.
\end{equation}
For gene G2, there are 5 states: G2-unbound or free,
G2$p_1$-bound by $p_1$, G2P-bound by RNAP, G2$p_1p_2$-bound by
$p_1$ and $p_2$ and G2$p_1P$-bound by $p_1$ and RNAP, which leads
to,
\begin{equation}
[G2] + [G2p_1] +[G2P] + [G1p_1P] + [G1p_1p_2]=1.
\end{equation}
The following reaction scheme defines the network,
\medskip
\begin{table}
\begin{center}
\begin{tabular}{ll} \hline\\
$p_1 + G1$ & $\rightleftharpoons  ~~~~G1p_1$~~~~( $k_{1f}, k_{1b}$) \\
$P + G1$ & $\rightleftharpoons  ~~~~G1P$~~~~( $k_{1pf}, k_{1pb}$) \\
$P + G1p_1$ & $\rightleftharpoons  ~~~~G1p_1P$~~~~( $k_{2pf}, k_{2pb}$) \\
$G1P $ & $\rightarrow  ~~~~G1P + p_2$~~~~( $k_{p_2}$)\\
$G1p_1P $ & $\rightarrow  ~~~~G1p_1P + p_2$~~~~( $k_{p_2}$)\\
$p_2 $ & $\rightarrow  ~~~~\phi$~~~~( $\gamma_{2}$)\\
$p_1 + G2$ & $ \rightleftharpoons ~~~~ G2p_1$~~~~( $k_{ 3f}, k_{3b}$) \\
$G2p_1 + p_2$ & $\rightleftharpoons  ~~~~G2p_1p_2 $~~~~( $k_{4f}, k_{4b}$)\\
$P + G2p_1$ & $ \rightleftharpoons ~~~~ G2p_1P$~~~~( $k_{3pf}, k_{3pb}$) \\
$P + G2$ & $\rightleftharpoons  ~~~~G2P$~~~~( $k_{4pf}, k_{4pb}$) \\
$G2P $ & $\rightarrow  ~~~~G2P + p_3$~~~~( $k_{p_3}$)\\
$G2p_1P $ & $\rightarrow  ~~~~G2p_1P + p_3$~~~~( $k_{p_4}$)\\
$p_3 $ & $ \rightarrow ~~~~ \phi$~~~~( $\gamma_3$)\\
\\\hline
\end{tabular}
\end{center}
\caption{Reaction Scheme for the feedforward network.}
\end{table}
\medskip
From the above reaction scheme, we obtain at thermodynamic
equilibrium,
\begin{eqnarray}
[G1p_1]&=&k_1~[p_1]~[G1], [G1p_1P]=k_{2p}~[G1p_1]~[P],
[G1P]=k_{1p}~[G1]~[P]\\ \nonumber [G2p_1]&=&k_{3}~[G2]~[p_1],
[G2p_1p_2]=k_{4}~[G2p_1]~[p_2], [G2P]=k_{4p}~[G2]~[P],\\ \nonumber
[G2p_1P]&=&k_{4p}~[G2p_1]~[P]\\ \nonumber
\end{eqnarray}
where in the above equations we use the equilibrium constants,
which are ratios of the forward to backward rates, {\it e.g} $k_1
= \frac{k_{1f}}{k_{1b}}$, {\it etc}. Using Eqns. 17, 18 \& 19, the
ratio of genes bound by RNAP, and hence the transcription rates
can be evaluated to be,
\begin{eqnarray}
Tr_{G1}&\propto&\frac{k_{1p}~P +k_1~k_{2p} p_1~P}{1+k_1~p_1+k_1~k_{2p}~p_1~P+k_{1p}~P}\\
\nonumber Tr_{G2}&\propto&\frac{k_{3p}~P +k_2~k_{4p} p_1~P}{1+k_{3p}~P+k_2~p_1+k_1~k_{3}~p_1~p_2+k_2~k_{4p}~p_1~P}\\
\nonumber
\end{eqnarray}
which is functionally the same in form to Eq. 17. Both methods are
equivalent as the only physical requirement is thermal equilibrium
for these reactions. Although the statistical approach is more
intuitive, the equilibrium approach allows us to define a reaction
scheme, which can be described in terms of measurable kinetic
constants.

\section{Appendix 2}

In each of the subsections below, the transcriptional rates for
the genes, which are described by the truth table for the
transcriptional rules, are derived. The parameter values used for
the simulations are also provided. The parameters used in the equations are
lumped, in the sense that we group together terms which depend
on the same variable, {\it e.g}
$\alpha_3~[p_1]~[P]+\alpha_2~[p_1]=v_1~p_1$ {\it etc}. We also
assume for simplicity that $P=1$.

\subsection{Feedforward Networks}

\subsubsection{Simple Feedforward}

Table 1, in Appendix I describes the regulation at G1 and G2. The
parameter values used in Eq. 4 are,
\begin{table}[h]
\begin{center}
\begin{tabular}{|c|c|c|c|c|c|c|c|c|c|c|}
\hline
$a_1$ & $a_2$ & $a_3$ & $a_4$  &  $\gamma_1$  & $b_1$& $b_2$  &  $b_3$ &  $b_4$ &  $b_5$ &$\gamma_2$\\
\hline
0&0.01&0&0.002&0.01&0&1&0&0.011&3~$10^{-4}$&0.08\\
\hline
\end{tabular}
\caption{Parameters values for Eq. 4.}
\end{center}
\end{table}

\subsubsection{Mixed Feedforward}
\begin{table}[htb]
\centering
\parbox[t]{0.45\textwidth}{
\centering
\begin{tabular}{ccr}
\hline
$p_1$ & $P$ & Rate \\
\hline \hline
0 & 0 & $1$ \\
1 & 0 & $\delta_1 [p_1]$ \\
0 & 1 & $\delta_2 [P]$ \\
1 & 1 & $\delta_3 [p_1~P]$ \\
\hline
\end{tabular}
}
\parbox[t]{0.45\textwidth}{
\centering
\begin{tabular}{cccr}
\hline
$p_1$ & $p_2$ & $P$  & Rate \\
\hline \hline
0 & 0 & 0 & $1$ \\
0 & 1 & 0 & $\epsilon_1 [p_2]$ \\
1 & 0 & 0 & $\epsilon_2 [p_1]$ \\
1 & 1 & 0 & $\epsilon_3 [p_1][p_2]$ \\
0 & 0 & 1 & $\epsilon_4 [P]$ \\
0 & 1 & 1 & $\epsilon_5 [p_2][P]$\\
1 & 0 & 1 & $\times$ \\
1 & 1 & 1 & $\times$ \\
\hline
\end{tabular}
} \caption{Logic for the transcriptional regulation underlying
mixed feedforward network genes, Eq. 7. }
\end{table}
\begin{eqnarray}
Tr_{G1}&\propto&\frac{\delta_2~[P] +
\delta_3~[p_1]~[P]}{1+\delta_2~[P]+\delta_1~[p_1]+\delta_3~[p_1]~[P]}\\
\nonumber Tr_{G2}&\propto&\frac{\epsilon_4~[P] +
\epsilon_5~[p_2]~[P]}{1+\epsilon_4~[P]+\epsilon_1~[p_2]+\epsilon_2~[p_1]+\epsilon_3~[p_1]~[p_2]+\epsilon_5~[p_2]~[P]}\\ \nonumber
\end{eqnarray}
\begin{table}[h]
\begin{center}
\begin{tabular}{|c|c|c|c|c|c|c|c|c|c|c|c|c|}
\hline
$c_1$ & $c_2$ & $c_3$ & $c_4$  &  $\gamma_c$  & $\gamma_1$&$d_1$& $d_2$  &  $d_3$ &  $d_4$ &  $d_5$ &$d_6$&$\gamma_2$\\
\hline
0&0.01&0&0.002&0.0001&0.01&0&0.5&0&0.001&0.0051&$10^{-4}$&0.09\\
\hline
\end{tabular}
\caption{Parameters values for Eq. 7.}
\end{center}
\end{table}

\subsubsection{Consequence of Dimerization for a Simple Feedforward and the effects of Mutations at
G1, G2.}

\begin{table}[h]
\begin{center}
\begin{tabular}{|c|c|c|c|c|c|c|c|c|c|c|c|c|c|c|}
\hline
$e_1$ & $e_2$ & $e_3$ & $e_4$  &  $\gamma_1$ &$k_{d1}$&$k_{d2}$&$k_{d3}$&$d_1$& $d_2$  &  $d_3$ &  $d_4$ &  $d_5$&$\gamma_2$\\
\hline
0&0.01&0&0.002&0.01&20&2000&0.001&0&1&0&.011&0.003&0.08\\
\hline
\end{tabular}
\caption{Parameters values for Eq. 6.}
\end{center}
\end{table}
The upper plot in Fig 4, for the simple feedforward (monomer) we
used the same parameters values as in Table3. For the lower two
figures, the mutations were carried on the parameters $e_2, d_5$.
Their values are given by, panel A: $e_2=0.001$-wild type,
$e_2=2\times10^{-4}$-mutant, panel B:$d_5=0.003$-wild type,
$d_5=5\times10^{-5}$-mutant

\subsubsection{Application 1: Time Ordering}

The amplitude filter parameters used for the simulations in Fig. 6
are the same as in Table 3, except for the following changes given
by,
\begin{table}[h]
\begin{center}
\begin{tabular}{|c|c|c|c|}
\hline
Filter \#& $e_2$ & $d_5$ & $\gamma_2$ \\
\hline
1&0.001&0.05&0.07\\
\hline
 2 & $5~10^{-4}$ & $5~10^{-4}$ & 0.3 \\
\hline
\end{tabular}
\caption{Parameters values for Fig. 6.}
\end{center}
\end{table}

\subsubsection{Application 2: Homeostatic Networks}

The regulation at G1 in Fig. 7 (a) is assumed to be,
\begin{table}[htb]
\centering
\parbox[t]{0.45\textwidth}{
\centering
\begin{tabular}{cccr}
\hline
$A$ & $p_3$ & $P$& Rate \\
\hline \hline
0 & 0 & 0&$1$ \\
1 & 0 & 0&$\rho_1 [A]$ \\
0 & 1 &0 &$\rho_2 [p_3]$ \\
1 & 1 &0 &$\rho_3 [A][p_3]$ \\
0 & 0 & 1&$\rho_4 [P]$ \\
1 & 0 & 1&$\times$ \\
0 & 1 &1&$\times$ \\
1 & 1 &1 &$\rho_7 [A][p_3][P]$ \\
\hline
\end{tabular}
} \caption{Logic for the transcriptional regulation underlying G1
for the homeostatic network. }
\end{table}
\begin{equation}
Tr_{G1}\propto \frac{\rho_4 [P]+\rho_7 [A][p_3][P]}{1+\rho_1 [A]+\rho_2 [p_3]+\rho_3 [A][p_3]+ \rho_4 [P]+\rho_7 [A][p_3][P] }
\end{equation}
The parameter values used in Eq. 8, to generate Fig. 8 are given
by,
\begin{table}[h]
\begin{center}
\begin{tabular}{|c|c|c|c|c|c|c|}
\hline
$u_1$ & $u_2$ & $u_3$ & $u_4$ & $u_5$ & $u_6$ & $\gamma_1$ \\
\hline
0&100&0&0.01&0.001&11&0.1\\
\hline
\end{tabular}
\caption{Parameters values for Eq. 7.}
\end{center}
\end{table}
and the parameters used for the amplitude filter are the same as
in Table 4.

\subsection{Tunable Motifs}

\begin{table}[htb]
\centering
\parbox[t]{0.4\textwidth}{
\centering
\begin{tabular}{cccccr}
\hline
$A$ & $p_1$ & ${p_2}^4$ & $p_3$ & $P$& Rate \\
\hline \hline
0 & 0 & 0 &0 & 0&$1$ \\
1 & 0 & 0 &0 & 0&$\varepsilon_1[A]$ \\
0 & 0 & 1 &0 & 0&$\varepsilon_2[{p_2}^4]$\\
1 & 0 & 1 &0 & 0&$\varepsilon_3[A][{p_2}^4]$ \\
0 & 0 & 0 &1 & 0&$\varepsilon_4[p_3]$ \\
0 & 0 & 1 &1 & 0&$\varepsilon_5 [{p_2}^4][p_3] $ \\
1 & 0 & 0 &0 & 1&$\varepsilon_6 [A][P] $ \\
0 & 0 & 0 &1 & 1&$\varepsilon_7 [p_3][P] $ \\
\hline
\end{tabular}
} \centering
\parbox[t]{0.2\textwidth}{
\centering
\begin{tabular}{ccr}
\hline
${p_1}^4$ & $P$ & Rate \\
\hline \hline
0 & 0 & $1$ \\
1 & 0 & $\zeta_1 [{p_1}^4]$ \\
0 & 1 & $\zeta_2 [P]$ \\
1 & 1 & $\times$ \\
\hline
\end{tabular}
}
\parbox[t]{0.2\textwidth}{
\centering
\begin{tabular}{ccr}
\hline
$p_2$ & $P$  & Rate \\
\hline \hline
0 & 0 & $1$ \\
1 & 0 & $\eta_1 [p_1]$ \\
0 & 1 & $\eta_2 [P]$ \\
1 & 1 & $\eta_3 [p_1][P]$ \\
\hline
\end{tabular}
} \caption{Logic for the transcriptional regulation underlying the
Tunable Motif network genes, for G1, G2 and G3 respectively. }
\end{table}
\begin{eqnarray}
Tr_{G1}&\propto&\frac{\varepsilon_6 [A][P]+\varepsilon_7 [p_3][P]}{1+\varepsilon_1[A]+\varepsilon_2[{p_2}^4]+\varepsilon_3[A][{p_2}^4]+\varepsilon_4[p_3] +\varepsilon_5 [{p_2}^4][p_3]+\varepsilon_6 [A][P]+\varepsilon_7 [p_3][P]}\\
\nonumber Tr_{G2}&\propto&\frac{\zeta_2 [P]}{1+\zeta_2 [P]+ \zeta_1 [{p_1}^4]}\\
\nonumber
Tr_{G3}&\propto&\frac{\eta_2 [P]+\eta_3 [p_2] [P]}{1+ \eta_1 [p_2]+\eta_2 [P]+\eta_3 [p_2] [P]}\\
\nonumber
\end{eqnarray}
The parameter values in Eq. 8, are given by,
\begin{table}[h]
\begin{center}
\begin{tabular}{|c|c|c|c|c|c|c|c|c|}
\hline
$m_1$ & $m_2$ & $m_3$ & $m_4$  & $m_5$& $m_6$ &$m_7$&$\gamma_1$&$n_1$\\
\hline $2~10^{-3}$&0.2&0.01&1&0.01&0.01&0.01&0.1&0.15\\
\hline $n_2$&$n_3$& $\gamma_2$  &  $o_1$ &  $o_2$ &  $o_3$&$o_4$&$\gamma_2$ & \\
\hline 0.001&1&.1&0&0.01&0&0.001&0.01 &\\
\hline
\end{tabular}
\caption{Parameters values for Eq. 8.}
\end{center}
\end{table}

\subsection{Adjustable Gates}

\begin{table}[htb]
\centering
\parbox[t]{0.45\textwidth}{
\centering
\begin{tabular}{ccccr}
\hline
$p_1$ & $p_2$ & $p_3$ & $P$& Rate \\
\hline \hline
0 & 0 & 0 &0 & $1$ \\
1 & 0 & 0 &0 & $\theta_1 [p_1]$ \\
0 & 1 & 0 &0 & $\theta_2 [p_2]$\\
0 & 0 & 1 &0 & $\theta_3 [p_3]$\\
1 & 0 & 1 &0 & $\theta_4 [p_1][p_3]$\\
0 & 1 & 1 &0 & $\theta_5 [p_2][p_3]$\\
1 & 1 & 0 &0 & $\theta_6 [p_1][p_2]$\\
1 & 1 & 1 &0 & $\theta_7 [p_1][p_2][p_3]$\\
1 & 0 & 1 &1 & $\theta_8 [p_1][p_3][P]$\\
0 & 1 & 1 &1 & $\theta_9 [p_2][p_3][P]$\\
1 & 1 & 0 &1 & $\theta_{10} [p_1][p_2][P]$\\
\hline
\end{tabular}
} \centering
\parbox[t]{0.45\textwidth}{
\centering
\begin{tabular}{ccccr}
\hline
$p_1$ & $p_2$ & $p_3$ & $P$& Rate \\
\hline \hline
0 & 0 & 0 &0 & $1$ \\
1 & 0 & 0 &0 & $\vartheta_1 [p_1]$ \\
0 & 1 & 0 &0 & $\vartheta_2 [p_2]$\\
0 & 0 & 1 &0 & $\vartheta_3 [p_3]$\\
1 & 0 & 1 &0 & $\vartheta_4 [p_1][p_3]$\\
0 & 1 & 1 &0 & $\vartheta_5 [p_2][p_3]$\\
1 & 1 & 0 &0 & $\vartheta_6 [p_1][p_2]$\\
1 & 1 & 1 &0 & $\vartheta_7 [p_1][p_2][p_3]$\\
1 & 0 & 0 &1 & $\vartheta_8 [p_1][P]$\\
0 & 1 & 0 &1 & $\vartheta_9 [p_2][P]$\\
1 & 1 & 0 &1 & $\vartheta_{10} [p_1][p_2][P]$\\
1 & 1 & 1 &1 & $\vartheta_{11} [p_1][p_2][p_3][P]$\\
\hline
\end{tabular}
} \caption{Logic for the transcriptional regulation underlying the
adjustable gates, {\it i.e.} regulation of G1, for the two cases
considered in the main text (Eq. 12 \& Eq. 15 respectively). The
Table on the left refers to the case for which G1 behaves as an
{\bf AND} gate for low $p_3$, and an {\bf OR} gate for high $p_3$.
The table on the right is for the opposite case.}
\end{table}
\begin{eqnarray}
Tr_{G1}^{Adjustgate1}&\propto& \frac{z_{on1}}{z_{on1}+z_{off1}},
\\ \nonumber
z_{on1}&=&  \theta_8 [p_1][p_3][P]+\theta_9
[p_2][p_3][P]+\theta_{10} [p_1][p_2][P] ,\\ \nonumber z_{off1}&=&
1+\theta_1 [p_1]+\theta_2 [p_2]+\theta_3 [p_3]+\theta_4
[p_1][p_3]+\theta_5 [p_2][p_3]+\theta_6 [p_1][p_2]+\theta_7
[p_1][p_2][p_3],\\ \nonumber
Tr_{G1}^{Adjustgate2}&\propto&\frac{z_{on2}}{z_{on2}+z_{off2}},
\\ \nonumber
z_{on2}&=& \vartheta_8 [p_1][P]+\vartheta_9
[p_2][P]+\vartheta_{10} [p_1][p_2][P]+\vartheta_{11}
[p_1][p_2][p_3][P], \\ \nonumber z_{off2}&=& 1+ \vartheta_1
[p_1]+\vartheta_2 [p_2]+\vartheta_3 [p_3]+\vartheta_4
[p_1][p_3]+\vartheta_5 [p_2][p_3]+\vartheta_6
[p_1][p_2]+\vartheta_7 [p_1][p_2][p_3].\\ \nonumber
\end{eqnarray}
\begin{table}[htb]
\centering
\parbox[t]{0.45\textwidth}{
\centering
\begin{tabular}{ccr}
\hline
$p$ & $P$& Rate \\
\hline \hline
0 & 0 & $1$ \\
1 & 0 & $\psi_1 [T]$ \\
0 & 1 & $\psi_2 [P]$\\
0 & 0 & $\psi_3 [T][P]$\\
 \hline
\end{tabular}
} \centering
\parbox[t]{0.45\textwidth}{
\centering
\begin{tabular}{ccccr}
\hline
$[p_4]^4$ & $P$& Rate \\
\hline \hline
0 & 0 & $1$ \\
1 & 0 & $\chi_1 [p_4]^4$ \\
0 & 1 & $\chi_2 [P]$\\
0 & 0 & $\chi_3 [p_4]^4 [P]$\\
\hline
\end{tabular}
} \caption{Logic for the transcriptional regulation for genes G2
\& G3 in Fig. 14.}
\end{table}
\begin{eqnarray}
Tr_{G2}&=& \frac{\psi_2 [P]+\psi_3 [T][P]}{1+\psi_1 [T]+\psi_2 [P]+\psi_3 [T][P] }\\
\nonumber Tr_{G3}&=&\frac{\chi_2 [P]+\chi_3 [p_4]^4 [P]}{1+\chi_1
[p_4]^4+\chi_2 [P]+\chi_3 [p_4]^4 [P]},
\\ \nonumber
\end{eqnarray}
\begin{table}[h]
\begin{center}
\begin{tabular}{|c|c|c|c|c|c|c|c|c|c|c|}
\hline
$r_1$ & $r_2$ & $r_3$ & $r_4$  & $r_5$& $r_6$ &$r_7$&$r_8$&$r_9$& $r_{10}$\\
\hline
$4~10^{-4}$ & $4~10^{-4}$ & $4~10^{-4}$ &v0.01&0.01&0.01&0.01&0.01&0.01& $10^{-4}$  \\
\hline
$k_1$&$k_2$&$\gamma_p$  & $c_0$ &$\gamma_{p_4}$&$v_1$&$v_2$&$v_3$&$v_4$&$\gamma_{p_3}$\\
\hline
1&.056&0.04&0.1&0.001&0&0.001&0&0.001&0.01\\
\hline
\end{tabular}
\caption{Parameters values for Eq. 15.}
\end{center}
\end{table}

\end{bmcformat}
\end{document}